\documentclass[aps,pre,twocolumn,showpacs,superscriptaddress]{revtex4}
\usepackage{graphicx} 
\usepackage[dvips,
            colorlinks=true, 
            citecolor=blue, 
            urlcolor=blue,
            linkcolor=blue,
            ]{hyperref}
\usepackage{amssymb}%, amsmath}
\usepackage{bbm}
\begin{document}

\title{Signatures of fractal clustering of aerosols advected under gravity}

\author{Rafael D. Vilela}
%\email{rdvilela@pks.mpg.de} 
\affiliation{Max Planck Institute for the Physics of Complex Systems,
%N\"othnitzer Strasse 38, 
01187 Dresden, Germany}
 
\author{Tam\'as T\'el}
\affiliation{Institute for Theoretical Physics, 
E\"otv\"os University - P.O. Box 32, H-1518, Budapest, Hungary}

\author{Alessandro P.S. de Moura}
\affiliation{Department of Physics, King's College, University
of Aberdeen - Aberdeen AB24 3UE, UK}

\author{Celso Grebogi}
\affiliation{Institute for Complex Systems, King's College, University
of Aberdeen - Aberdeen AB24 3UE, UK}

%\thanks{}%
%\subjclass{}%
%\keywords{}%

%\date{}%
%\dedicatory{}%
%\commby{}%
% ----------------------------------------------------------------
\begin{abstract}
%Finite-size heavy particles 
Aerosols under chaotic advection often approach 
a strange attractor. They move chaotically on this
fractal set but, in the presence of gravity, they have a net
vertical motion downwards.
In practical situations,
observational data may be 
available only at a given level, for example at the ground level. 
We uncover two fractal signatures of chaotic advection 
of aerosols under the action of gravity.
Each one enables the computation of the fractal 
dimension $D_{0}$ of the strange
attractor governing the advection dynamics
from data obtained solely at a given level. 
We illustrate our theoretical
findings with a numerical experiment
and discuss their possible relevance to meteorology.
\end{abstract}
\maketitle

% ----------------------------------------------------------------

The transport of finite-size particles plays important role 
in several fields, 
from cloud physics \cite{malinowski} to plankton dynamics \cite{plankt}.
The recent interest \cite{fsp_strange_attractors,bec}
in this problem comes in part from the fact that 
the dynamics of these particles is dissipative 
due to the drag force.
This makes dynamical systems tools and concepts, such as attractors and
dimensions, applicable.
In the presence of gravity, 
a net vertical motion occurs due to the density difference between
fluid and particle. 
For heavy particles ({\it aerosols}), this leads 
to raindrop falling in the atmosphere \cite{prup} and to
the sedimentation of plankton \cite{plankt} and 
marine snow \cite{marine_snow} in the ocean.

In such situations,
knowledge of the advection dynamics of the aerosols 
is of fundamental importance,
whereas usually only data obtained at a given level (height) are available.
This occurs often, for instance,  in meteorology, 
in the case where the aerosols are raindrops.
In fact, it is much easier to obtain direct measurements of the 
raindrops when they reach the ground level
than before, i.e., when they are being advected in the air flow.
The derivation of approaches to obtain information on the 
advection dynamics of aerosols 
{\it solely} from data observed at a given level is, therefore, 
an instrumental and relevant task.
In particular, here we are interested in 
approaches to obtain the {\it fractal dimension}
of the set where the aerosols cluster while they are advected.

We report the uncovering of two independent 
fractal signatures of chaotic advection under gravity.
Both  
make 
the computation of the fractal 
dimension $D_{0}$ of the strange
attractor in the 
$N$-dimensional configuration space possible 
without prior knowledge of the
advection dynamics.
First, we show that the time series of the instants
of arrival of advected aerosols in a small detector placed at 
a given level
has a fractal dimension which is equal to
\begin{equation}
d_0=1+D_0-N.
\label{d0}
\end{equation}
We assume that $D_0 <N$, 
which implies that the attractor in the full $2N$-dimensional 
phase space is $D_0$-dimensional since a set of dimension $D_0$, 
when projected into a space of dimension $N$, 
typically remains $D_0$-dimensional if $D_0 <N$ \cite{Falconer}. 
Second, we show that
the spatial distribution 
of the aerosols reaching 
a line at a given level
contains discontinuities ({\it jumps}) at points that form
a fractal set whose dimension is again equal to
$d_0$ \cite{obs}.  
We illustrate our findings with a
numerical experiment.

The dimensionless form of the governing equation for
the path ${\bf{r}}(t)$ of aerosols {\it much} denser
than the fluid, subjected to Stokes drag and gravity, 
reads as \cite{maxey_riley}:
\begin{equation}
\ddot{\bf r}=A\left({\bf u}-\dot{\bf r}-W{\bf n}\right),
\label{maxey}
\end{equation}
where $\dot{\bf r}$ is the velocity of the aerosol,
${\bf u}={\bf u}({\bf r}(t),t)$ is the fluid velocity field
evaluated at the position ${\bf r}(t)$ of the aerosol, and
${\bf n}$ is a unit vector pointing upwards in the vertical 
direction. Throughout this paper
we consider the vertical direction along the axis $y$. 
The inertia parameter $A$ (larger values for smaller inertia)
can be written in terms  of the densities  $\rho_p$
and $\rho_f$
of the aerosol and of the fluid, respectively,
the radius $a$ of the aerosols, the fluid's
kinematic viscosity $\nu$, and the characteristic
length $L$ and velocity $U$ of the flow.
It is $A={R}/{St}$, where
$R={\rho_f}/{\rho_p}\ll 1$
and $St=(2a^2U)/(9\nu L)$ is the Stokes's number of the aerosol.
As seen from (\ref{maxey}),
the gravitational parameter $W$ provides the dimensionless settling
velocity in a medium at rest. The actual settling
velocity is the result of two effects:
the gravitational attraction (buoyancy)
and an updraft, if present.
We consider the settling velocity
to be comparable with $U$, implying a $W$ of the order of unity.

For convenience, we treat the case where the fluid
flow is two-dimensional, $N=2$.
In this situation, the phase space of the 
advection dynamics of the
aerosols is 4-dimensional,
since the aerosols are not constrained to move with
the same velocity as their corresponding fluid elements.
For the sake of concreteness, let us
consider the
time-smoothened version of the
alternating sinusoidal shear flow of Ref. \cite{pierrehumbert}.
In dimensionless form it is given as
\begin{equation}
\left\{ \begin{array}{l}
u_x({\bf r},t)=0.5\left(1+\tanh(\gamma\sin(2\pi t))\right)\sin(2\pi y),  \\[0.3cm]
u_y({\bf r},t)=0.5(1-\tanh(\gamma\sin(2\pi t)))\sin(2\pi x). \label{fluid}
\end{array} \right.
\end{equation}
This flow is defined on the unit square with periodic
boundary conditions, and is periodic in time with a unit period.
The vertical direction corresponds to the $y$-axis.
Apart from the spatial (sinusoidal) factor, each velocity component
consists of two plateaus in time with a rapid but smooth crossover if
$\gamma=20/\pi$.
This is a simple analytically given model which, nevertheless, possesses a
paradigmatic property  of both  chaotic and turbulent flows
\cite{Vil}:
the intense stretching of material elements.

Substituting Eq. (\ref{fluid}) into Eq. (\ref{maxey})
and fixing $R=10^{-3}$ and
$W=0.8$,
there are
regions of the parameter $St$ for which
the dynamics of the aerosols is ruled by a strange attractor.
We note that the time-independence of the attractor's dimension
$D_0$, which will be essential in what follows, is  valid for a broad
class of randomly time-dependent flows as well (see conclusion).
A bifurcation diagram and our analysis 
suggest the existence of strange attractors
of dimension between 1 and 2 in an interval around
$St=2\times 10^{-4}$ 
(apart, as usual, from some small periodic windows).
Here we illustrate our general ideas by analyzing
the strange attractor corresponding to $St=2\times 10^{-4}$, 
which is prototypical.
Figure 1 shows four snapshots of the projection, $\Lambda$, of the 
strange attractor into the configuration space.
Because of gravity, 
there is a net vertical motion of the fractal set of curves 
of $\Lambda$ downwards in Fig. 1.
We compute the dimension of $\Lambda$ to be $D_0=1.79$.

\begin{figure}
%\centerline{\includegraphics[angle=0,width=7cm,height=6.0cm]{figure1.eps}} 
\includegraphics[width=0.45\textwidth]{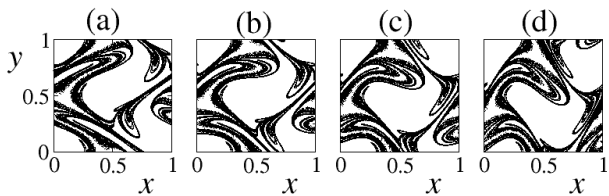}
\caption{
Snapshots, taken at (a) $t=0.65$, (b) $0.75$, (c) $0.85$ and (d) $0.95$,  
of the projection $\Lambda$ of 
the strange attractor into the configuration 
space. 
}
\label{fig1} 
\end{figure}

{\it Arrival times--} 
The first fractal signature can be inferred from the following
argument: 
Let us fix a point $P={\bf r}_{0}$ in the configuration space. 
Consider the product space of the configuration space
and the time axis. In this space-time representation, 
the point $P$ corresponds to a $1$-dimensional 
straight line $({\bf r}_{0},t)$, $t \in \mathbb{R}$.
The time coordinates of the set formed by the 
intersection of this line with the 
$(D_0 +1)$-dimensional time-extension
of $\Lambda$ correspond to the time instants of arrivals of
aerosols at $P$.
This intersection has a dimension \cite{Falconer} given by
$d_0 = 1 + (D_0+1) - (N+1)$, 
since the time-extended configuration space is $(N +1)$-dimensional.
Therefore, the time series 
of the instants
of arrival of advected aerosols at the point $P$ 
has a fractal dimension $d_0$ given by (\ref{d0}).

For practical purposes, we substitute the point $P$ by a 
small codimension 1 object, placed horizontally,  
which is our {\it detector}. In the two-dimensional
fluid flow that we analyze here, such object is a segment of 
length $\delta\ll L$.
Seen at the scale of the detector, the dynamics of the 
advected aerosols 
under gravity roughly
corresponds to a Cantor set of curves moving with some
(both time and space dependent) horizontal velocity component $v_h$.
The fractality of the 
time series of the instants of arrival of aerosols at the 
detector is measurable at scales $\tau>{\delta}/{v_h}$.

To validate numerically this fractal signature, 
we let $10^6$ aerosols, distributed on the attractor within
the unit square at time zero, evolve 
until they reach the level  $y=y_0=0$. 
Their instants of arrival as well as their $x$ coordinates 
are then recorded.
We then choose a point $x_0$ and analyze the time series of instants of 
arrival at the segment $x\in (x_0,x_0+\delta)$, $y=0$.
The dimension $d_{ts}$ of this time series should, in face of our 
theoretical findings, be equal to $d_0=0.79$.
We find excellent agreement with this value. 
For instance, taking $\delta=10^{-3}$, 
for $x_0=0.2$, $x_0=0.5$ and $x_0=0.8$, we 
find, respectively, $d_{ts}=0.80$, $d_{ts}=0.77$ and $d_{ts}=0.80$.
The numbers of points in each of these choices of $x_0$
are, respectively, 185, 663 and 562.
To calculate these dimensions, 
we have used the method described in \cite{grassberger_tel}. 
This method is much more efficient
than the usual box counting  and is specially 
suited for computing the dimension of subsets of the real line.
We explain it briefly here:
Let us say that we want to measure the dimension $d$ of
the set $M$. For each point $b_i \in M$, let $n_i(l)$
be the number of points in $M$ that lie within a distance $l$
of $b_i$. The following scaling can be shown to hold:
$\langle 1/n(l) \rangle \sim {l^{-d}}$, 
where $d$ is the fractal dimension, and the bracket denotes 
a uniform averaging over all elements of the set.  
The dimensions for the different choices of $x_0$ were measured 
with $l \equiv \tau$ in the range $0.002<\tau<0.05$, equally spaced
on the logarithmic scale.   
 
{\it Density jumps--}
The second fractal signature is associated with
the spatial distribution $\mathbb{P}$ of the
advected aerosols reaching 
a line $s$ at 
a certain level $y=y_0$
of the configuration space. 
Let ${\bf r }_{0}  \in \mathbb{R}^{N-1}$
be the spatial coordinate of a point along a line $s$ 
at the 
$(N-1)$-dimensional plane $y=y_0$.
We define 
$\mathbb{P}({\bf r}_0)\mbox{d}{\bf r}$ as 
the probability that an aerosol reaches
this plane  
in a  $\mbox{d}{\bf r}$-neighborhood
of the point ${\bf r}_0$ over a finite time interval $\Delta t$. 
$\mathbb{P}({\bf r}_0)$ is proportional to 
the frequency of particles falling near ${\bf r}_0$,
and can in principle be determined experimentally. The idea of the
second signature is that the downward motion with local velocity ${\bf v}$ 
caused by gravity is
equivalent to a projection of the whole fractal pattern shown in
Figs. 1, 3(a), and 4 onto a horizontal line. As illustrated in Fig. 2, the
direction of the projection is tangent to the distribution at a fractal
set of points, which causes discontinuous jumps in the projected measure
if $\Delta t$ is sufficiently large. 
This means that there is a fractal set of points where the density of
detected particles has discontinuities. This is explained more
rigorously in what follows, and we show that the fractal dimension of
$\Lambda$ can be obtained from the dimension of the set of points
where $\mathbb{P}$ changes discontinuously.

The attractor has an $SRB$-like measure \cite{srb}, which
is absolutely continuous (discontinuous) along the unstable (stable) foliation, and
the distribution of points on $\Lambda$
inherits this property. 
The distribution 
$\mathbb{P}$ is proportional to the projected natural measure
in the box of size $\mbox{d}{\bf r}$ around  ${\bf r}_0$ 
on the level $y=y_0$
integrated over 
time. 
The measure is not continuous in the configuration space, 
but rather concentrated on the filaments
of $\Lambda$. As a result, the distribution $\mathbb{P}$ 
can have a discontinuous jump
where the local velocity ${\bf v}$ at a point of $\Lambda$ happens to be tangent
to the corresponding filament. 
We call the points where this happens
{\it extremal points} \cite{observation}.
In order to determine the dimension of such points, 
let us first  consider a plane $(x,y)$ of the projected attractor 
(see Fig. 2). 
The extremal points of the main filaments can be joined by a 
smooth line $\xi$ (Fig. 2). The attractor's dimension on this plane is 
$D_0+2-N$ (it is $D_0$ and
$D_0-1$ for $N=2$ and $N=3$, respectively).   
Thus, the dimension of the extremal points on this plane is that
of the intersection of a $1$ and a  $D_0+2-N$-dimensional object, which is
$1+(D_0+2-N)-2=D_0 + 1 -N$. 
Next, observe that in the product space of the $x,y$ plane and the time axis
this is a $D_0+2-N$-dimensional set.
The points of line $s$ ($y=y_0, z=$const) on which the distribution 
$\mathbb{P}$ is defined
form a plane in this product space. Thus the set of points where the density jumps
occur {\em in the line $s$ at the fixed level $y=y_0$} 
has a dimension $2+ (D_0+2-N) -3= D_0+1-N=d_0$.

\begin{figure}
%\centerline{\includegraphics[angle=0,width=7cm,height=2.8cm]{third_figure3.eps}} 
\includegraphics[width=0.45\textwidth]{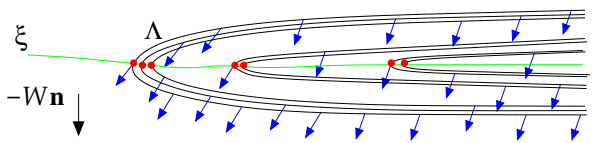}
\caption{(color online)
Illustration justifying the fractal signature
on the density jumps. 
The black curves represent the slice $z=$const of 
$\Lambda$ 
at a certain time instant $t$. 
The vector field ${\bf v}$ defined on $\Lambda$
is represented by the blue arrows.
Its tangencies with $\Lambda$ are the extremal points (red dots).
The line $\xi$ (green) is a 1-dimensional line which joins all the extremal 
points of the $(x,y)$ plane. 
}
\label{fig2} 
\end{figure}

The second fractal signature refers to a {\it local} property
in the sense that it is related to discontinuities
of a distribution. 
It is, therefore, intrinsically 
more difficult to detect in an experiment,
due to 
the fluctuations induced 
by the finite number of observed aerosols.
Such fluctuations tend to obscure the true discontinuities
and lead
to the consideration of points which do not 
correspond to true discontinuities.
In principle, it can, however, be detected
for a sufficiently large number of aerosols.
To show this,  
consider  
the part of the attractor
which is shown in black in Fig. 3(a).
We let $1.7\times 10^{6}$ aerosols, forming that part of 
the attractor, evolve
until they reach the line $y=y_0=0$.
We measure the spatial distribution of the
aerosols here using a sliding-window method, to be described now. 
We consider  
the segment 
$(0.88,0.98)\times 0$, which is divided into $10^5$
equal boxes $s_i$. We count the number $n_i$ of aerosols arriving in each
$s_i$. We then define $r_i =\sum_{j=i}^{i+100} n_j$,
$i=1, \cdots, 10^5-100$, the total  number of arrived aerosols
in $100$ neighboring boxes. Next, consider
$\Delta_i=r_{i+100}-r_i$, $i=1, \cdots, 10^5-200$,
the  
jumps between two adjacent intervals, each formed
by $100$ boxes.
Figure 3(b) shows $\Delta_i$
in the whole range. 
The local maxima are also indicated.
We define a threshold $\alpha$ for these  jumps
and consider that the discontinuities
in the spatial distribution occur at
the points 
corresponding to the local maxima of $\Delta_i$
which exceed $\alpha$.
The procedure is robust for $\alpha$ in a certain range that, 
at the same time, allows the detection of a reasonably large
number of discontinuities and does not indicate false discontinuities
(due to the fluctuations).
In our numerical experiment, the minimum value of $\alpha$
for which no false discontinuity is detected is $204$.
In this case, we find $51$ discontinuities.
We compute the dimension $d_{sd}$
of this set of discontinuities.
We obtain 
$d_{sd}=0.82$ over the range $0.0004<l<0.01$
in the interval $0.88 < x < 0.98$. 
The subset of $\Lambda$ reaching the line
$y=0$ at the points where these discontinuities occur
can be seen in Fig. 4.
For $\alpha=218$, we find $47$ discontinuities, 
and the corresponding dimension is $d_{sd}=0.81$ over the same range.
These dimensions are 
in very good agreement with the expected value $d_0=0.79$.
For these computations,  we use again the method of  
Ref.~\cite{grassberger_tel}. 

\begin{figure}
%\centerline{\includegraphics[angle=0,width=8.4cm,height=3.8cm]{figure3new.eps}}
\includegraphics[width=0.45\textwidth]{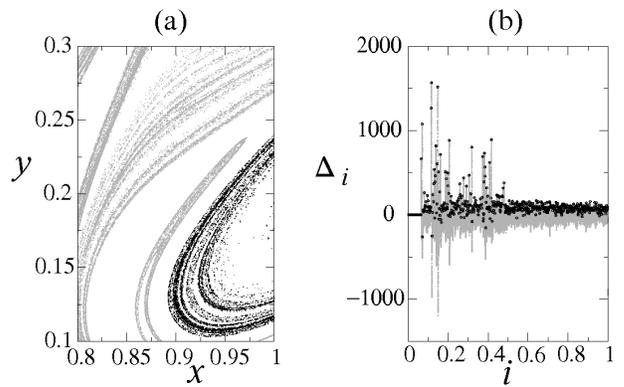}
\caption{
(a) Magnification of the snapshot $t=0.85$ of $\Lambda$
from Fig.1.
Both the black and the gray points belong to $\Lambda$.
The black points correspond to the part of the attractor which
is used
to illustrate the second fractal signature. 
(b) Density jumps $\Delta_i$ as a function
of the box index $i$ (in units of $10^5$) (gray)
along the line $s$ given by $y=y_0=0$. 
%(the box size is $\delta=10^{-6}$).
The local maxima are also shown (black dots). 
}
\label{fig3} 
\end{figure}

\vspace{2mm}
 
\begin{figure}
\includegraphics[width=0.45\textwidth]{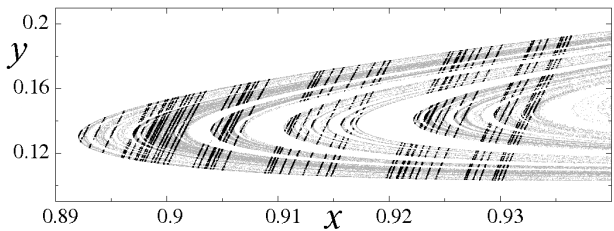}
\caption{
Branch of $\Lambda$ at $t=0.85$ (gray and black points)
from which
the spatial distribution of aerosols is computed 
in the level $y=y_0=0$.
The black subset reaches the level $y=0$ 
at the points where the density jumps ($\Delta_i>\alpha$)
occur.
}
\label{fig4} 
\end{figure}

To summarize, we have shown that important information of
the advection dynamics of aerosols in fractal sets can be
obtained from data measured solely at a given level. 
We have uncovered two
independent fractal signatures of chaotic advection under gravity.
We have illustrated our fractal signatures using a flow
model with periodic time-dependence, but our findings
are far more general, since flows with {\it random} 
time-dependence (yet spatially smooth) also have the properties
mentioned here. In particular, aerosols advected in such flows
often approach a random attractor characterized by
a well defined time-independent fractal dimension 
\cite{Rom,SO,bec}.
Note that the weight of the particles does not play a role in the argument.
Therefore, the fractal signatures derived would also apply for the rising dynamics
of finite-size particles lighter than the ambient fluid ({\it bubbles}).
In fact, our fractal signatures are applicable to fractal chaotic attractors in general.

Our findings might be useful in meteorology, in the case 
where the aerosols are raindrops.
Although our chaotic advection model 
neglects possibly important
features of rain precipitation such as 
the spatial roughness of turbulent flows 
and the distribution of raindrop sizes, 
the fractal signature 
on the arrival times relies {\it only} on the time invariance
of the dimension of the set where the aerosols accumulate.
Therefore, if either the aforementioned mechanism based on the
convergence of aerosols in random flows
to strange attractors \cite{obs4}
or {\it any} other mechanism leads to the accumulation 
of raindrops in fractal sets with a time-invariant dimension \cite{obs5},
then the signature on the arrival times 
may be used to characterize the fractal clustering of the raindrops.
This signature
could be measurable in precipitation data of
rain, for instance, with the disdrometer
which was used in the experiment reported in \cite{lavergnat}.
We note that a disdrometer was already used in Ref. \cite{zawadzki}
and even the {\it correlation dimension} ($D_2$) of a time
series was measured and interpreted as a
sign of irregular distribution of drops in space.
Finally, we mention 
the experimental results of
Ref. \cite{lavergnat} 
showing a fractal dimension for 
the time series of arrival times of raindrops at a disdrometer
(Fig. 9(a) of the cited reference).
Both results \cite{lavergnat,zawadzki}
are compatible with 
the accumulation of raindrops in fractal sets with a time-invariant dimension.

%\begin{acknowledgments}
The authors thank I. Geresdi, 
I.J. Benczik, J. Davoudi, S.P. Malinowski, K. Gelfert, 
and an anonymous referee for useful discussions and suggestions.
The support of FAPESP and CNPq (Brazil) and 
of the Hungarian Science Foundation 
(OTKA T047233, TS044839) are acknowledged.
This work was partially done at Instituto de F\'\i sica, 
Universidade de S\~ao Paulo, S\~ao Paulo, Brazil.
%\end{acknowledgments}


\begin{thebibliography}{99}
\bibitem{obs}
If the dimension of the strange attractor is 
smaller than $N-1$, then Eq. (\ref{d0}) yields
$d_0 < 0$, implying that both the set of time
instants of arrival of aerosols in a small detector
and the set
where discontinuities occur in the spatial
distribution along a line at the fixed level are empty.
\bibitem{observation}
Whether an extremal point is actually 
associated with a density jump depends
on whether it belongs or not to the 
line $s$ at the level $y=y_0$.
\bibitem{obs4}
The parameters %$R,St$ and $W$ 
in our numerical experiment have 
been chosen to match raindrop data.
For raindrops, 
typical numbers are $R \approx 10^{-3}$ and  
$a\approx 1$mm. 
%Using chaotic advection  
Considering the advection of these drops by
%chaotic-advection-described 
coherent structures of linear size $L\approx100$m
and typical velocity fluctuations
$U\approx1$m/s in atmospheric flows, 
we obtain $St \approx 2 \times 10^{-4}$ 
and $A \approx 5$, 
since $\nu \approx 10^{-5}m^2/s$ for air.
\end{thebibliography}
\end{document}